\documentclass[twocolumn]{aastex631}

\begin{document}

\title{A multi-wavelength investigation of spiral structures in $z > 1$ galaxies with JWST}

\correspondingauthor{Boris S. Kalita}
\email{boris.kalita@ipmu.jp; kalita.boris.sindhu@gmail.com}

\author[0000-0001-9215-7053]{Boris S. Kalita}
\altaffiliation{Kavli Astrophysics Fellow}
\affiliation{Kavli IPMU (WPI), UTIAS, The University of Tokyo, Kashiwa, Chiba 277-8583, Japan}
\affiliation{Kavli Institute for Astronomy and Astrophysics, Peking University, Beijing 100871, People's Republic of China}
\affiliation{Center for Data-Driven Discovery, Kavli IPMU (WPI), UTIAS, The University of Tokyo, Kashiwa, Chiba 277-8583, Japan}

\author[0000-0002-3462-4175]{Si-Yue Yu}
\affiliation{Kavli IPMU (WPI), UTIAS, The University of Tokyo, Kashiwa, Chiba 277-8583, Japan}

\author[0000-0002-0000-6977]{John D. Silverman}
\affiliation{Kavli IPMU (WPI), UTIAS, The University of Tokyo, Kashiwa, Chiba 277-8583, Japan}
\affiliation{Department of Astronomy, School of Science, The University of Tokyo, 7-3-1 Hongo, Bunkyo, Tokyo 113-0033, Japan}
\affiliation{Center for Astrophysical Sciences, Department of Physics \& Astronomy, Johns Hopkins University, Baltimore, MD 21218, USA}

\author[0000-0002-3331-9590]{Emanuele Daddi}
\affiliation{CEA, Irfu, DAp, AIM, Universit\`e Paris-Saclay, Universit\`e de Paris, CNRS, F-91191 Gif-sur-Yvette, France}

\author[0000-0001-6947-5846]{Luis C. Ho}
\affiliation{Kavli Institute for Astronomy and Astrophysics, Peking University, Beijing 100871, People's Republic of China}
\affiliation{Department of Astronomy, School of Physics, Peking University, Beijing 100871, People's Republic of China}

\author[0000-0002-9382-9832]{Andreas L. Faisst}
\affiliation{Caltech/IPAC, MS 314-6, 1200 E. California Blvd. Pasadena, CA 91125, USA}

\author[0000-0003-0348-2917]{Miroslava Dessauges-Zavadsky}
\affiliation{Department of Astronomy, University of Geneva, 51 Chemin Pegasi, 1290 Versoix, Switzerland}

\author[0000-0001-9369-1805]{Annagrazia Puglisi}
\altaffiliation{Anniversary Fellow}
\affiliation{School of Physics and Astronomy, University of Southampton, Highfield SO17 1BJ, UK}

\author[0000-0003-3195-5507]{Simon Birrer}
\affiliation{Department of Physics and Astronomy, Stony Brook University, Stony Brook, NY 11794, USA}

\author[0000-0001-9044-1747]{Daichi Kashino}
\affiliation{National Astronomical Observatory of Japan, 2-21-1 Osawa, Mitaka, Tokyo 181-8588, Japan}

\author[0000-0002-0786-7307]{Xuheng Ding}
\affiliation{School of Physics and Technology, Wuhan University, Wuhan 430072,  China}

\author[0000-0001-9187-3605]{Jeyhan S. Kartaltepe}
\affiliation{Laboratory for Multiwavelength Astrophysics, School of Physics and Astronomy, Rochester Institute of Technology, 84 Lomb Memorial Drive, Rochester, NY 14623, USA}

\author[0000-0002-9252-114X]{Zhaoxuan Liu}
\affiliation{Kavli IPMU (WPI), UTIAS, The University of Tokyo, Kashiwa, Chiba 277-8583, Japan}
\affiliation{Center for Data-Driven Discovery, Kavli IPMU (WPI), UTIAS, The University of Tokyo, Kashiwa, Chiba 277-8583, Japan}
\affiliation{Department of Astronomy, School of Science, The University of Tokyo, 7-3-1 Hongo, Bunkyo, Tokyo 113-0033, Japan}

\author[0000-0002-2603-2639]{Darshan Kakkad}
\affiliation{Centre for Astrophysics Research, University of Hertfordshire, College Lane, Hatfield AL10 9AB, UK}

\author[0000-0001-6477-4011]{Francesco Valentino}
\affiliation{Cosmic Dawn Center (DAWN), Denmark} 
\affiliation{Niels Bohr Institute, University of Copenhagen, Jagtvej 128, DK-2200 Copenhagen N, Denmark}

\author[0000-0002-7303-4397]{Olivier Ilbert}
\affiliation{Aix Marseille Univ, CNRS, CNES, LAM, Marseille, France}

\author[0000-0002-4872-2294]{Georgios Magdis}
\affiliation{Cosmic Dawn Center (DAWN), Denmark} 
\affiliation{DTU-Space, Technical University of Denmark, Elektrovej 327, 2800, Kgs. Lyngby, Denmark}
\affiliation{Niels Bohr Institute, University of Copenhagen, Jagtvej 128, DK-2200, Copenhagen, Denmark}

\author[0000-0002-7530-8857]{Arianna S. Long}
\affiliation{Department of Astronomy, The University of Washington, Seattle, WA 98195, USA, Austin, TX, USA}

\author[0000-0002-8412-7951]{Shuowen Jin}\altaffiliation{Marie Curie Fellow}
\affiliation{Cosmic Dawn Center (DAWN), Denmark}
\affiliation{DTU-Space, Technical University of Denmark, Elektrovej 327, 2800 Kgs. Lyngby, Denmark}

\author[0000-0002-6610-2048]{Anton M. Koekemoer}
\affiliation{Space Telescope Science Institute, 3700 San Martin Dr., Baltimore, MD 21218, USA} 

\author[0000-0002-6085-3780]{Richard Massey}
\affil{Department of Physics, Centre for Extragalactic Astronomy, Durham University, South Road, Durham DH1 3LE, UK}




\begin{abstract}
Recent JWST observations have revealed the prevalence of spiral structures at $z > 1$. Unlike in the local Universe, the origin and the consequence of spirals at this epoch remain unexplored. We use public JWST/NIRCam data from the COSMOS-Web survey to map spiral structures in eight massive ($> 10^{10.5}\,\rm M_{\odot}$) star-forming galaxies at $z_{\rm spec} \sim 1.5$. We present a method for systematically quantifying spiral arms at $z>1$, enabling direct measurements of flux distributions. Using rest-frame near-IR images, we construct morphological models accurately tracing spiral arms. We detect offsets ($\sim 0.2$ -- $0.8\,\rm kpc$) between the rest-frame optical and near-IR flux distributions across most arms. Drawing parallels to the local Universe, we conclude that these offsets reflect the presence of density waves. For nine out of eighteen arms, the offsets indicate spiral shocks triggered by density waves. Five arms have offsets in the opposite direction and are likely associated with tidal interactions. For the remaining cases with no detected offsets, we suggest that stochastic `clumpy' star formation is the primary driver of their formation. In conclusion, we find a multi-faceted nature of spiral arms at $z > 1$, similar to that in the local Universe.

\end{abstract}



\section{Introduction} \label{sec:intro}

Over $60\%$ of galaxies in the local Universe feature some level of spiral structure \citep{nair10a,willett13,buta15}. Nevertheless, the origin and consequences of spiral arms have been debated for nearly a century \citep[e.g.,][]{hubble26,reynolds27,vaucouleurs59,elmegreen82}. Given the current understanding \citep[see][for a review]{dobbs14, shu16}, the main lines of inquiry can be summarized into two key questions: (I) What are the \emph{formation mechanisms and physical characteristics of spiral arms}? (II) What is their \emph{impact on star formation} in their hosts? The answer to either question is far from straightforward. This paper focuses on the first question, aimed at the $z>1$ Universe, with the second to be explored in a follow-up work.

Decades of observations and simulations have established a few key modes of spiral arm formation: the quasi-stationary density wave theory, producing long-lived `grand design' spirals \citep{shu70,toomre77,shu16}; transient, recurrent spiral arms \citep{sellwood84,bottema03,fujii11,baba13}; local instability amplifications leading to ‘flocculent’ spirals \citep{mueller76,gerola78,elmegreen87}; and perturbations from tidal interactions \citep{toomre72,salo00b,dobbs10}. Emerging theories, such as manifolds \citep[for bar driven spirals; e.g.,][]{Athanassoula2009} and groove instability \citep{Sellwood2019}, also offer new interpretations of spiral arm formation.  Notably, these processes are not mutually exclusive, and can coexist within a galaxy \citep{elmegreen14, Yu2020}.

The quasi-stationary or transient density wave theory assumes a constant pattern speed that differs from the differentially rotating disk (except at the co-rotation radius, where both velocities match). Across most of the disk, within the co-rotation radius, the arms are expected to trail behind the disk. The kinematics of such a system is predicted to manifest as a color gradient across the width of the spiral arms \citep{gittins04,martinez09}. Some studies suggest these trends reflect stellar age gradients \citep{gonzalez96,martinez-garcia13}, while others argue that color gradients also arise from attenuation due to dust lanes along spiral arms \citep{yu18,Martinez-Garcia2023}, produced by spiral shocks \citep{lynds70,gittins04}. However, this evidence comes from only a few cases. Tidal interactions have also been suggested to produce density waves that trail the disk \citep{kalnajs73,binney08,oh08}, although strong perturbations could result in leading waves \citep{thomasson89,buta92,buta03}.

Meanwhile, some formation modes produce spiral arms that are expected to co-rotate with the disk. Stochastic star formation can lead to over-dense areas of stars, which are then sheared by the disk's differential rotation to form spirals \citep{mueller76,gerola78,sleath95,nomura01}. These local instability-driven flocculent spirals are not grand-design spirals but rather a patchwork of short, irregular arms. Tidal interactions can also create material arms that shear into co-rotating spirals \citep{toomre69,meidt13}. Thus, observational evidence of rotational velocity offsets has the potential to reveal the mechanisms behind spiral arm formation. However, the observational results discussed so far are limited to the local Universe.

\begin{figure*} 
    \centering
    \includegraphics[width=0.77\textwidth]{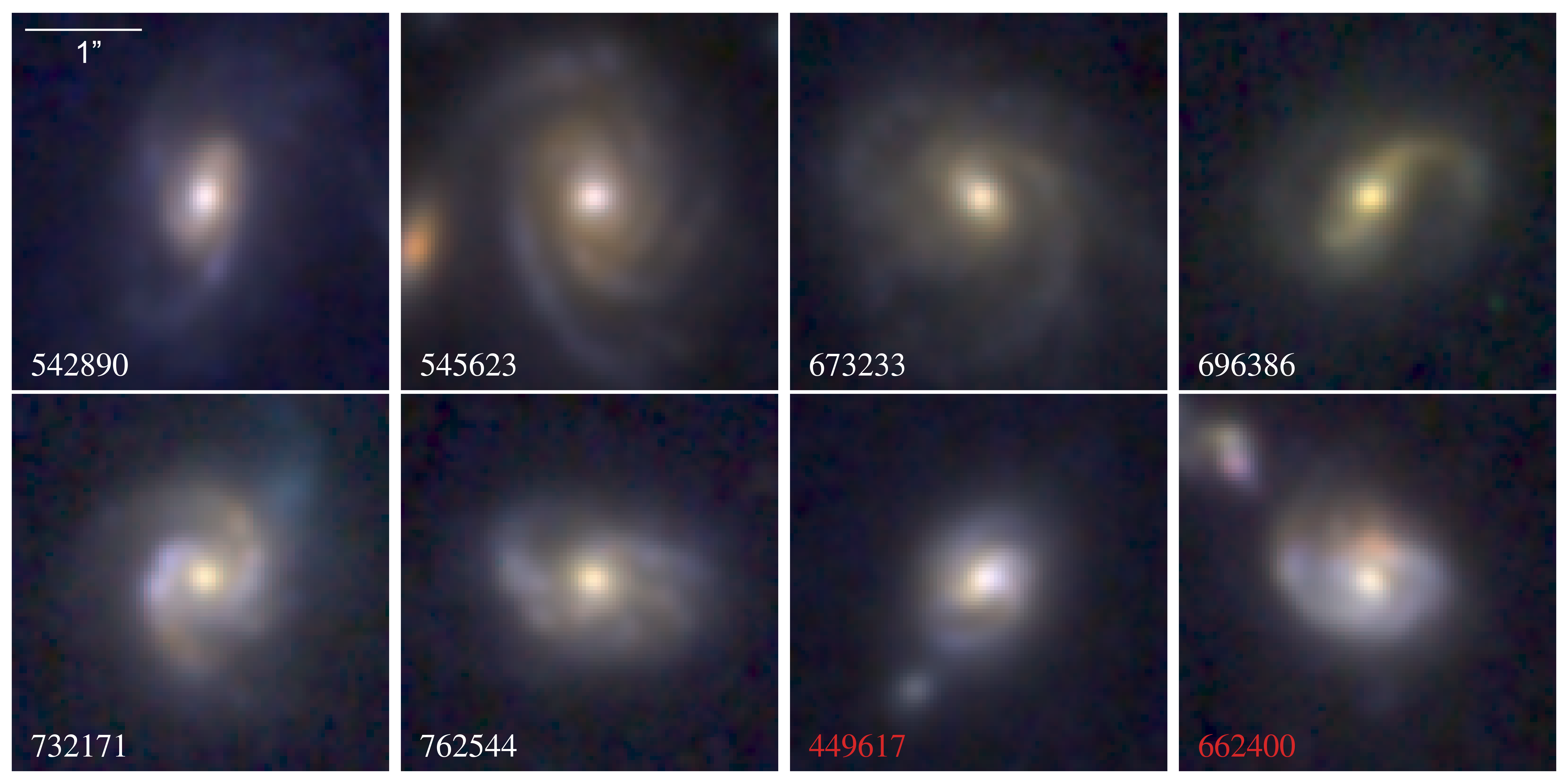}
    \includegraphics[width=0.77\textwidth]{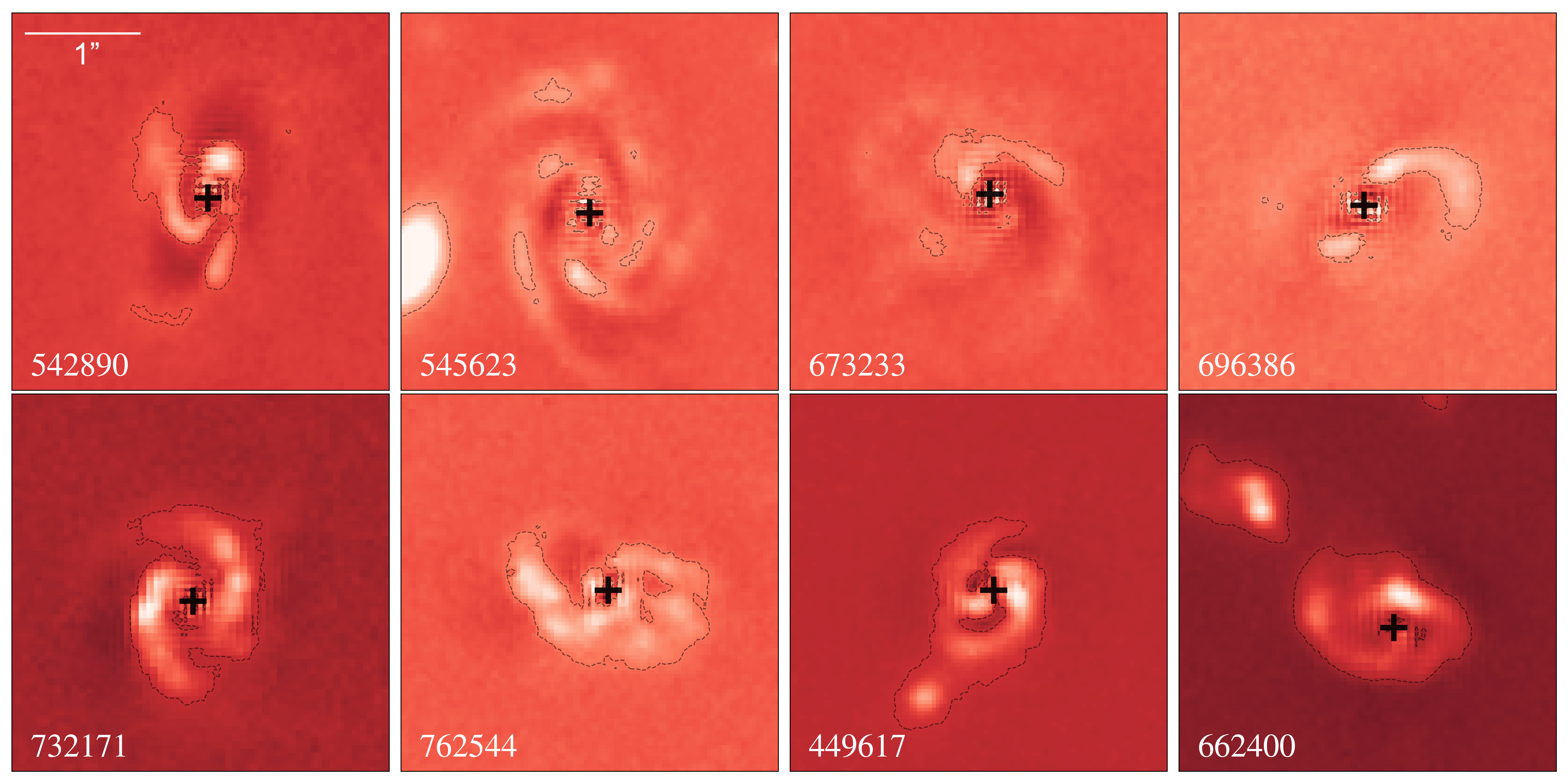}
    \includegraphics[width=0.77\textwidth]{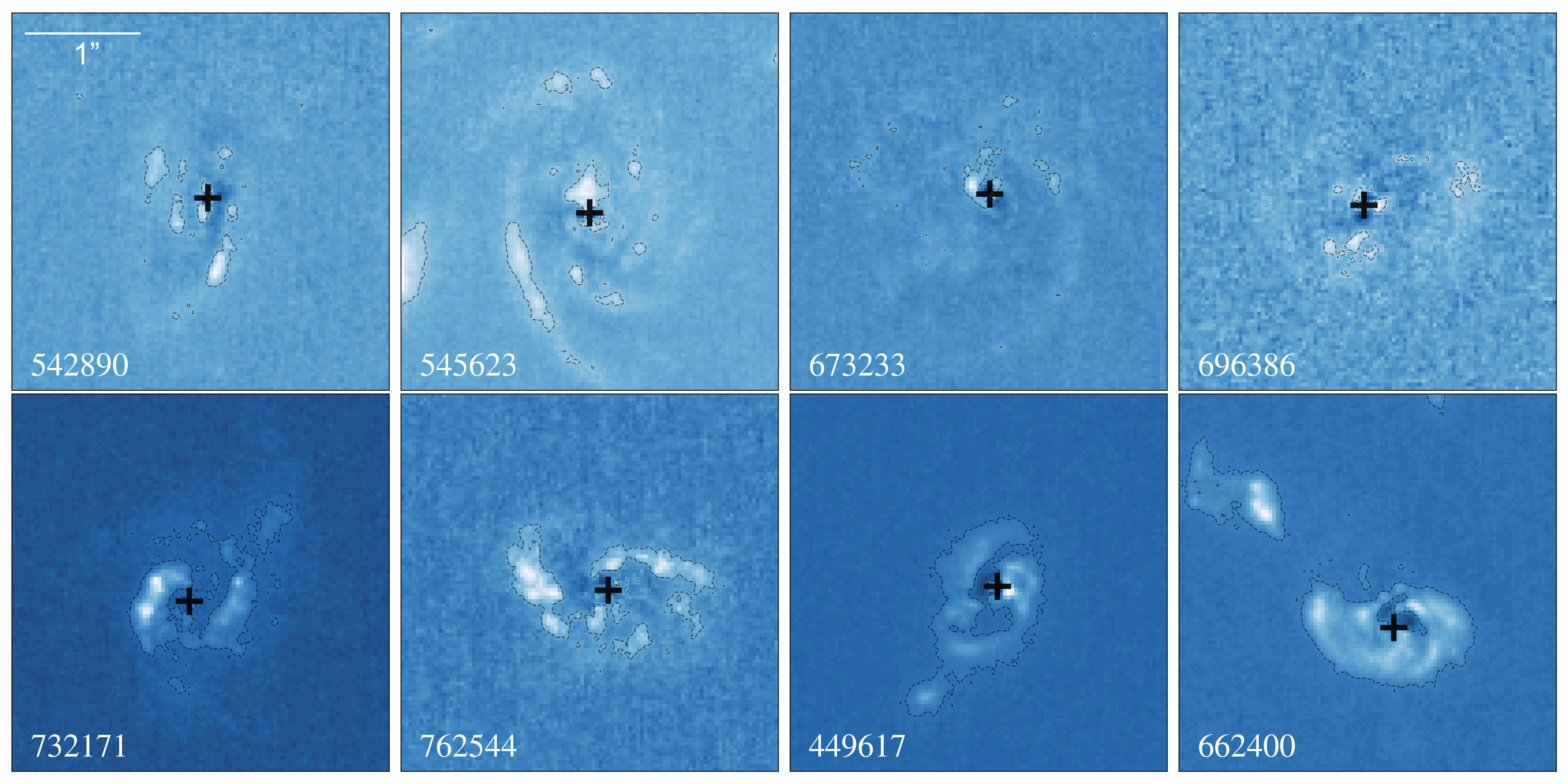}
    \caption{Image compilation of the sample: (top panels) The RGB images (F150W, F277W, F444W) of the 8 galaxies used in this work are $100 \times 100$ pixels, or $3^{\prime\prime} \times 3^{\prime\prime}$. The COSMOS2015 \citep{laigle16} IDs are provided for each galaxy, with the final two in red denoting the possible presence of an interacting companion. All are within a redshift range of $1.43 \leq z \leq 1.74$ and have stellar masses of $10^{10.5-11.4},\rm M_{\odot}$. The corresponding F444W (middle panels, with $5\sigma$ contours) and F150W (PSF-matched to F444W, bottom panels, with lower $3\sigma$ contours to account for the lower image depth) residual images after subtraction of a bulge and disk model highlight the presence of spiral arms.}
    \label{fig:image_comp}
\end{figure*}

Spiral arms in the $z>1$ regime have been relatively unexplored. The redshift range of $z = 1-3$ (Cosmic Noon) is, however, a critical phase of our Universe. In addition to marking the peak of the star formation rate density \citep{madau14}, significant morphological evolution of galaxies is expected \citep{toft14,lang14}. While discussions have largely focused on the build-up of central mass concentrations \citep[e.g.,][]{elbaz18,gomez-guijarro18,puglisi21,tan24,tan24b} and disk dynamics \citep{contini16,stott16,harrison17,posti18,marasco19,gillman20}, the evolution of key disk features like bars and spirals has received less attention \citep{elmegreen14,margalef-bentabol22,Martinez-Garcia2023, Yu2023, guo24}. Characterization of spirals at $z>1$ have mainly relied on visual classifications, which provide valuable insight into the variety of such features. 

With the advent of high-resolution JWST data revealing an abundance of spirals \citep[e.g.,][]{jacobs23, polletta24,mckinney24,liu24,kuhn24}, it is crucial to begin more detailed investigations of spirals during Cosmic Noon. This work initiates such an effort. Using JWST/NIRCam data for detailed modeling and flux distribution analysis, we aim to detect color gradients indicative of velocity offsets between spiral arms and the host disk. We construct the first method for a systematic quantification of spiral arms at $z>1$, and make direct flux distribution measurements across rest-frame optical and near-IR wavelengths. 

In this paper,  we introduce our sample in Sec.~\ref{sec:data}, followed by the analysis and results (Sec.~\ref{sec:analysis} and \ref{sec:results}, respectively). We conclude with a discussion (Sec.~\ref{sec:discussion}) and summary (Sec.~\ref{sec:summary}). Throughout, we adopt a concordance $\Lambda$CDM cosmology, characterized by  $\Omega_{m}=0.3$, $\Omega_{\Lambda}=0.7$, and $\rm H_{0}=70$ km s$^{-1}\rm Mpc^{-1}$. Magnitudes and colors are on the AB scale. All images are oriented such that north is up and east is left.



\section{Sample and Data} \label{sec:data}

We use eight galaxies (Fig.~\ref{fig:image_comp}, top panel) from the (H$\alpha$ detected) FMOS-COSMOS sample \citep{kashino13,silverman15,kashino19}, selected from 57 star-forming main-sequence galaxies within $\sim 0.3$ dex of the relation in \cite{speagle14} (FMOS-COSMOS-ALMA sample).  This sample has been presented in \cite{kalita24c}, and will be further discussed in upcoming works.  These galaxies are spectroscopically confirmed at $1.4 \leq z \leq 1.7$ with stellar masses of $10^{10.5-11.4}\,\rm M_{\odot}$ \citep[determined earlier in]{kashino19}. As they reside in the COSMOS field \citep{scoville07b}, 48 have multi-band (F115W, F150W, F277W, and F444W) JWST/NIRCam coverage from the COSMOS-Web survey \citep{casey23}.  After visually inspecting the images, we find eighteen of these galaxies clearly display spiral arms. Most of the rest also feature varying degrees of substructures, but they do not resemble spirals. Attempts at quantifying the visual classification will be addressed in upcoming works, since it requires a general estimate of the spiral strength across the full sample. Nevertheless, our `spiral-fraction' of $\sim 40\%$ is much higher than the $\lesssim 10\%$ expected at $z \sim 1.5$, found through visual inspections \citep{margalef-bentabol22}. We attribute this higher value to the likely bias introduced by the SFR-based selection of the sample. Of the eighteen galaxies, eight are found to be sufficiently uniform and bright (in near-IR) for our analysis, although two show signs of interaction with a companion (ids 449617 and 662400). Id 545623 also has a possible companion but features highly ordered spiral arms.

To map the spiral features in the rest-frame near-IR (tracing stellar mass distribution) and optical (representing unattenuated star formation), we use the F444W and F150W filters from the COSMOS-Web JWST/NIRCam dataset. The point-spread functions (PSFs) for each filter are created using the software \textsc{PSFEx} \citep{bertin11} on the full COSMOS-Web mosaic. Since dust likely plays a major role in determining flux distribution, we choose F444W over F277W to limit attenuation effects. F150W is preferred over F115W due to the deeper data ($5\sigma$ AB magnitude depths of 27.4 and 27.1, respectively) and a better sampled PSFs, as the image pixel scale ($0.03^{\prime\prime}$) in F115W is similar to its PSF FWHM ($0.04^{\prime\prime}$)\footnote{undersampled PSF can lead to systematic biases in spatial measurements.}. Finally, the F150W image is PSF-matched to the F444W data using a Gaussian kernel. 

\section{Analysis} \label{sec:analysis}
\subsection{Spirals in the near-IR residuals} \label{subsec:bulge-disk_residuals}
\begin{figure} 
    \centering
    \includegraphics[width=0.49\textwidth]{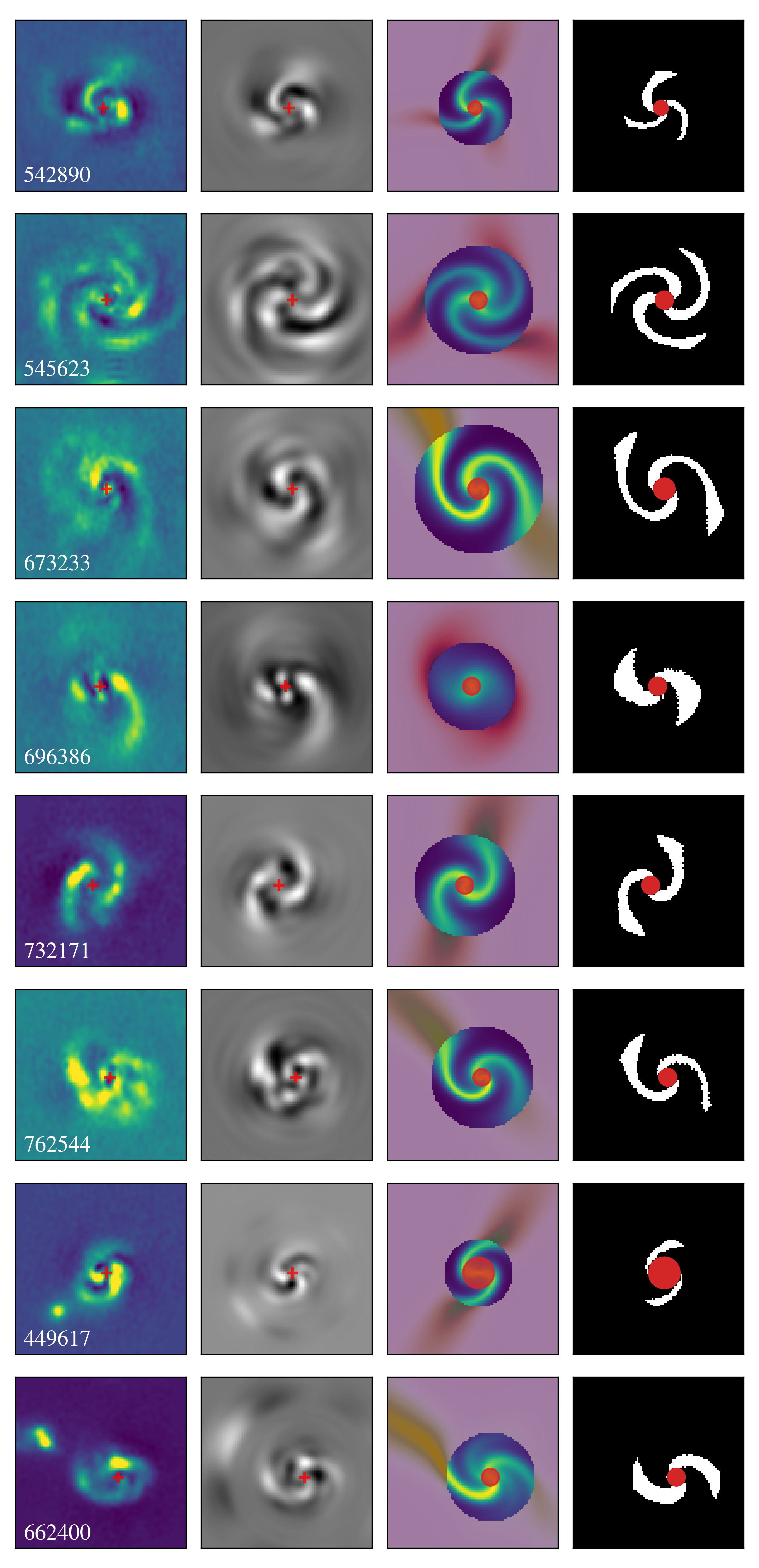}
    \caption{Locating the spiral arms: The de-projected residual F444W images (first column), the same image with only the m=$2-4$ components (second column), the GALFIT best-fit model for the disk+spiral components with the spiral arms amplified (as discussed in Sec.~\ref{subsec:spiral_models}; third column), and the segmentation regions created out of the model (fourth column) for each galaxy in our sample. The segmentation maps will be used to trace the location of each spiral arm. The red circles in the middle and right panels denote the bulge (effective radius) which has been excluded. The outer circle in the third panel shows the radial  limit of the analysis, which is twice the disk effective radius.}
    \label{fig:spiral_analysis}
\end{figure}

In this work, we aim to effectively model and quantify the spiral structures in the disk. The first step is to detect disk substructures, which is found to be most efficient by subtracting the underlying disk \citep[along with the bulge;][]{kalita24c}. Since the stellar distribution is best traced by the F444W rest-near-IR data, we model the corresponding images of our galaxy sample using Sérsic models. The fitting is performed with the Python-based package \textsc{GALIGHT}\footnote{https://github.com/dartoon/galight} \citep{ding20}, which implements the forward-modeling tool \textsc{LENSTRONOMY}\footnote{https://github.com/lenstronomy/lenstronomy} \citep{birrer18,birrer21}. This approach provides access to the full posterior distribution of each fitted parameter and is optimized using the Particle Swarm Optimizer \citep[PSO]{kennedy95}.

We compare the results of a single Sérsic model (without a fixed index) and a composite bulge-disk model (with Sérsic indices, $n = 2$ and $1$ for the bulge and disk, respectively)\footnote{We ensure that using a classical bulge with $n = 4$ does not change our results. However, we use the value for a pseudo-bulge \citep{fisher08} since the central region is star-forming throughout our sample}.  Based on the assessment of negative peaks in the residual images, reduced-$\chi^2$, and the Bayesian Information Criterion (BIC; which combines model complexity with $\chi^2$), we conclude that the bulge-disk model consistently outperforms the single Sérsic model, with the $-\Delta$BIC found to be $>10^3$. In the final residual near-IR images, after bulge-disk subtraction, the underlying spiral structures are clearly visible (Fig.~\ref{fig:image_comp}, middle panel). We do not refer to any specific flux or $\sigma$-threshold here since we already limit this study to galaxies where our spiral arm modeling method is successful. As mentioned in Sec.~\ref{sec:data}, a generalized determination of the flux values at which spiral arms can be successfully studied will be addressed in upcoming works. Finally, we create a residual optical (F150W) image by subtracting a bulge-disk model with shape parameters fixed to those from F444W and only fitting for flux. The resulting images are provided in Fig.~\ref{fig:image_comp} (bottom panel) and will be used in Sec.~\ref{subsec:mapping_flux}. 


\subsection{Determining spiral arm locations} \label{subsec:spiral_models}
Our goal is to measure the flux distribution over the spiral arms. To do so, we first need to locate the arms. This is typically achieved in the local Universe by performing discrete Fourier decomposition in polar coordinates, and then using the power spectrum to determine spiral arm characteristics \citep[see][for a discussion]{yu18b}. However, we find that such methods do not enable characterisation of spirals at $z>1$, due to the combined effects of weaker spiral arm strengths, irregularities, and additional substructures. Additionally, we find the arms become significantly more discontinuous in rest-frame optical wavelengths, further complicating characterisation.

The challenge of locating the arms directly from Fourier space information can be overcome by using shape priors in the image plane that are spiral models. Hence, we devise a modified method for locating spiral arms in our galaxy sample at $z \sim 1.5$. We first deproject the residual images onto a circular geometry using the axis ratio from the rest-frame near-IR disk fit (Sec.~\ref{subsec:bulge-disk_residuals}). This is done by rotating the image (to align the disk’s major axis with the x-axis) and decomposing it in the Cartesian coordinate system into shapelets using \textsc{LENSTRONOMY}. The image grid is then adjusted so the final axis ratio of the disk becomes 1. Since each galaxy in our sample already has an axis ratio $>0.7$, the change introduced is minimal.

The deprojected images (Fig.~\ref{fig:spiral_analysis}, first column) are then passed through a polar-shapelet transform (using \textsc{LENSTRONOMY}), where all azimuthal components except $m=2$, 3, and 4 are filtered out.  Higher order modes are sensitive to noise \citep{Kendall2011, Yu2021}, as well as `galaxy-clumps' expected at these redshifts \citep[e.g.,][]{genzel11,guo15,sattari23,claeyssens23,kalita24}. This step isolates structures that peak 2, 3, or 4 times per full rotation about the galaxy center (determined from the rest-frame near-IR bulge-disk model).  This method successfully extracts the spiral patterns as can be seen in Fig.~\ref{fig:spiral_analysis} (second column).   It is also noteworthy that the deprojection is essential for this step to work properly, as any residual circular component of the disk could be picked up as an $m=2$ component due to projection effects.

\begin{figure*} 
    \centering
    \includegraphics[width=0.77\textwidth]{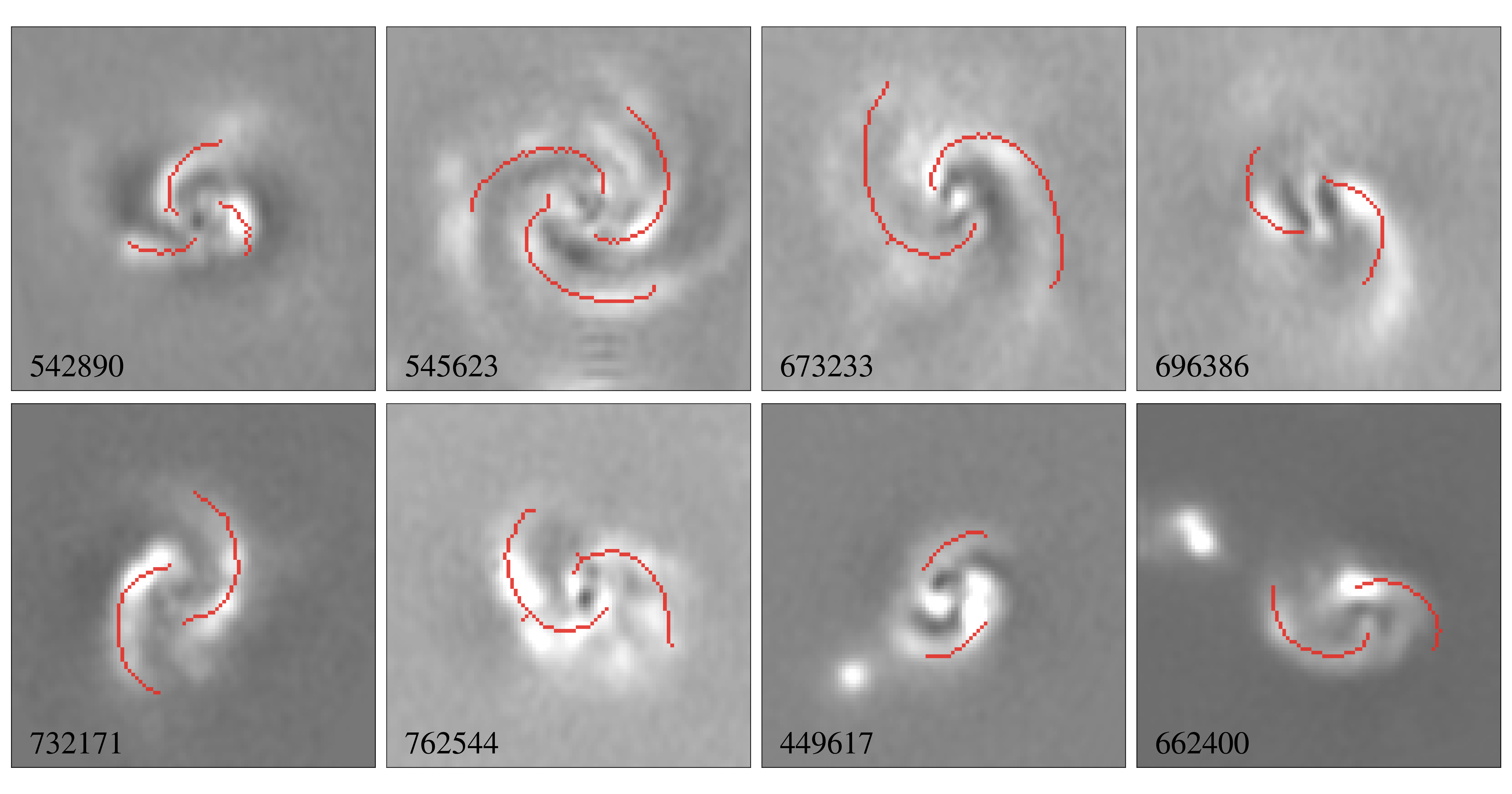}
    \caption{The model-based spiral skeletons overlaid on the de-projected residual F444W images of the galaxies in our sample.}
    \label{fig:spiral_skeletons}
\end{figure*}

The final part of the analysis aims to model the spiral structures.  To do so, we add back the now symmetric disk (deprojected version of the \textsc{GALIGHT} disk model with $n=1$) that was previously subtracted. This is essential since the spiral pattern on its own cannot be fit due to the presence of negative residuals. The final disk+spiral image is modeled using \textsc{GALFIT} \citep{peng10}, resulting in the final model of the spiral structures\footnote{We use a Sérsic model for the disk, with spiral sub-components with varying amplitudes, that are fit to the disk+spiral image}. Once fit, we artificially amplify the spiral arms by increasing the amplitude of the Fourier components of the GALFIT model (Fig.~\ref{fig:spiral_analysis}, third column). These models are used to generate segmentation maps (Fig.~\ref{fig:spiral_analysis}, fourth column), excluding the bulge (up to the effective radius in the rest-frame near-IR). The spiral arm paths (Fig.~\ref{fig:spiral_skeletons}) are then traced on the segmentation maps using the \textsc{skeletonize} function from the \textsc{skimage}\footnote{https://scikit-image.org/} Python package.

For the final part of the analysis, we use \textsc{GALFIT} because it includes the option to fit spiral components, a feature missing in \textsc{GALIGHT}. However, for the initial bulge+disk fitting, we prefer \textsc{GALIGHT}, as its PSO optimization outperforms the $\chi^{2}$-minimization approach of \textsc{GALFIT}, especially in the presence of strong substructures, which are common in our sample.  Due to the same reason, we are also unable to fit a bulge+disk+spiral model in \textsc{GALFIT}.

\begin{figure*} 
    \centering
    \includegraphics[width=0.46\textwidth]{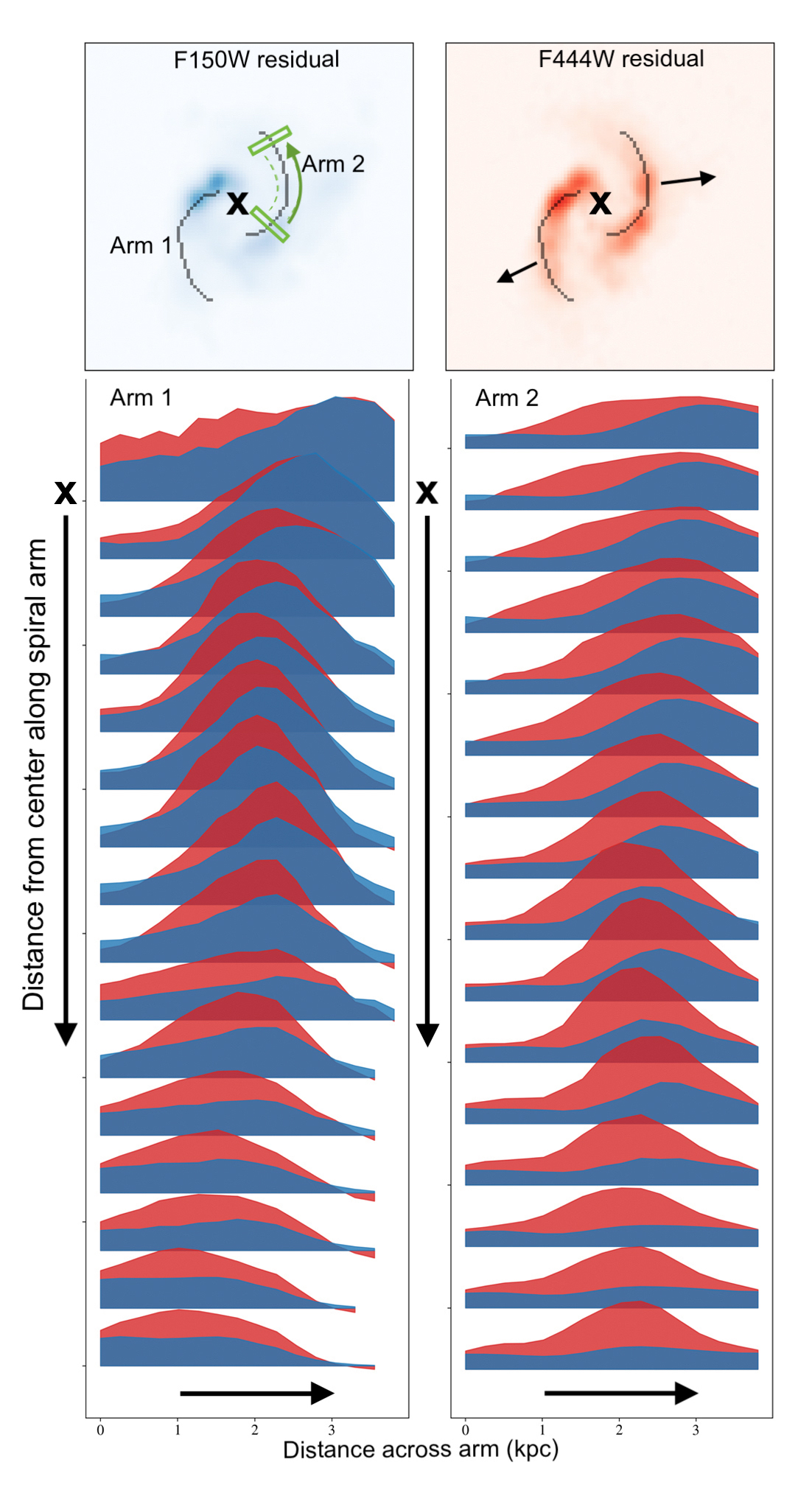}
    \includegraphics[width=0.46\textwidth]{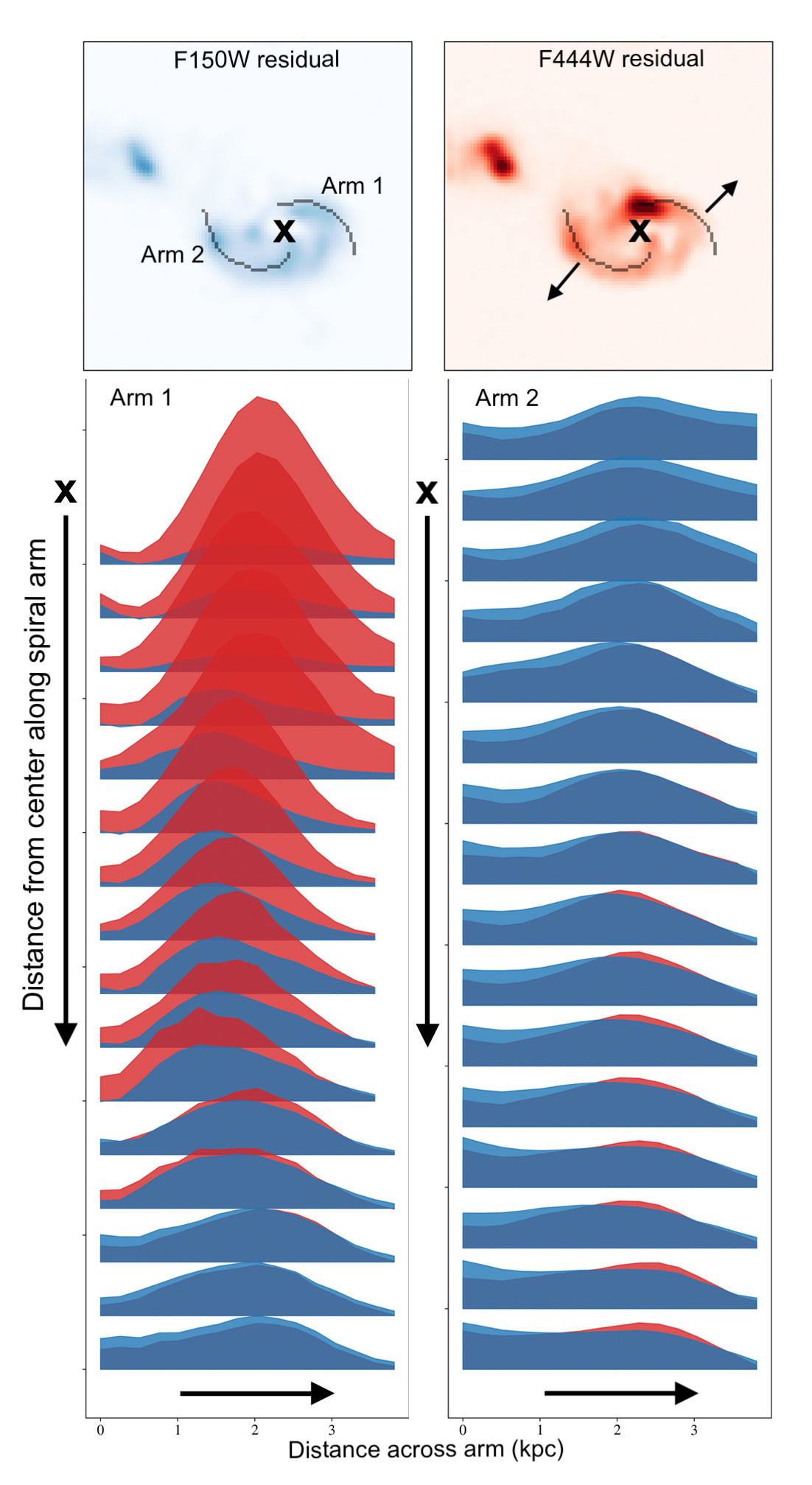}
    \caption{(Left)Positive color offsets in id 732171: (top panels) Residual F150W and F444W images after bulge and disk subtraction, with spiral arm paths over-plotted and expected propagation direction indicated by arrows. The first image on the top left also shows examples of the masks perpendicular to the arm used for flux measurements, with the longer edge being the length of the mask (lower panels) Flux distribution (all distributions at same flux scale) across each arm and along the length of the masks, with F150W in blue and F444W in red, for the first 16 pixels ($\sim 4\,\rm kpc$) along the arms starting from the center of the galaxy (marked with `x'). The distributions show clear positive offsets, with blue generally preceding red. The same arm propagation direction defines the x-axis (in kpc, calculated for $z=1.5$) ordering. The y-axis corresponds to the measurement bins along the arms, representing radial distance from the galaxy center in kpc. The first 16 iterations, with 2-pixel steps, are shown for each arm. (Right) Negative color offsets in id 662400: The same information, but for a case showing negative offsets. The blue F150W flux distribution generally trails behind the red F444W distribution.}
    \label{fig:pos_offset}
\end{figure*}

\subsection{Mapping flux across the arms} \label{subsec:mapping_flux}
From here onward, we only rely on the arm paths (Fig.~\ref{fig:spiral_skeletons}) determined in the previous section, and no longer use the GALFIT models. We measure the flux along the spiral arms (radially outward) in steps of 2 pixels, up to a radial distance of $2 \times$ the effective radius of the disk in the rest-frame near-IR. The per-pixel flux distribution is mapped at every step over a mask of 15 pixels in length ($\sim 4\,\rm kpc$ at $z=1.5$) and 3 pixels in width ($\sim 1.3 \times$ the full width half maximum of the F444W PSF). At each iteration, the radial direction of the arm is determined, and the mask length is kept perpendicular to it. Flux measurements are made for the deprojected residual images in both the rest-frame optical (F150W) and near-IR (F444W), mapping the variation along the mask's length while collapsing (averaging) over the width. Results for two of the galaxies are shown in Fig.~\ref{fig:pos_offset} (left and right). Also shown are two example masks, with the longer edge of each across the spiral arm being their lengths. 

We locate the peak of flux distribution across the spiral arms by fitting a skewed Gaussian model over the length of each mask. The positional error is given by the $1\sigma$ uncertainty of the fit. We check the centroid as well as the peak of the flux distributions, and find them to be in agreement with the model peak within the respective uncertainties. To estimate the systematic uncertainty of our procedure, we add an artificial point source to a source-free region of the data and perform the same deprojection procedure as in Sec.~\ref{subsec:spiral_models}. We then reverse the process, concluding that any systematic offset introduced in the first half would remain at the end of the full procedure. We estimate an uncertainty of $\sim 0.5$ pixel width ($\sim 0.13\,\rm kpc$ at $z=1.5$). The other source of uncertainty, due to intrinsic noise fluctuations, is already accounted for by this process. We exclude the polar-shapelet transform from this error estimate, as it is used only to determine mask position and orientation.

\section{Results} \label{sec:results}
\begin{figure*} 
    \centering
    \includegraphics[width=0.45\textwidth]{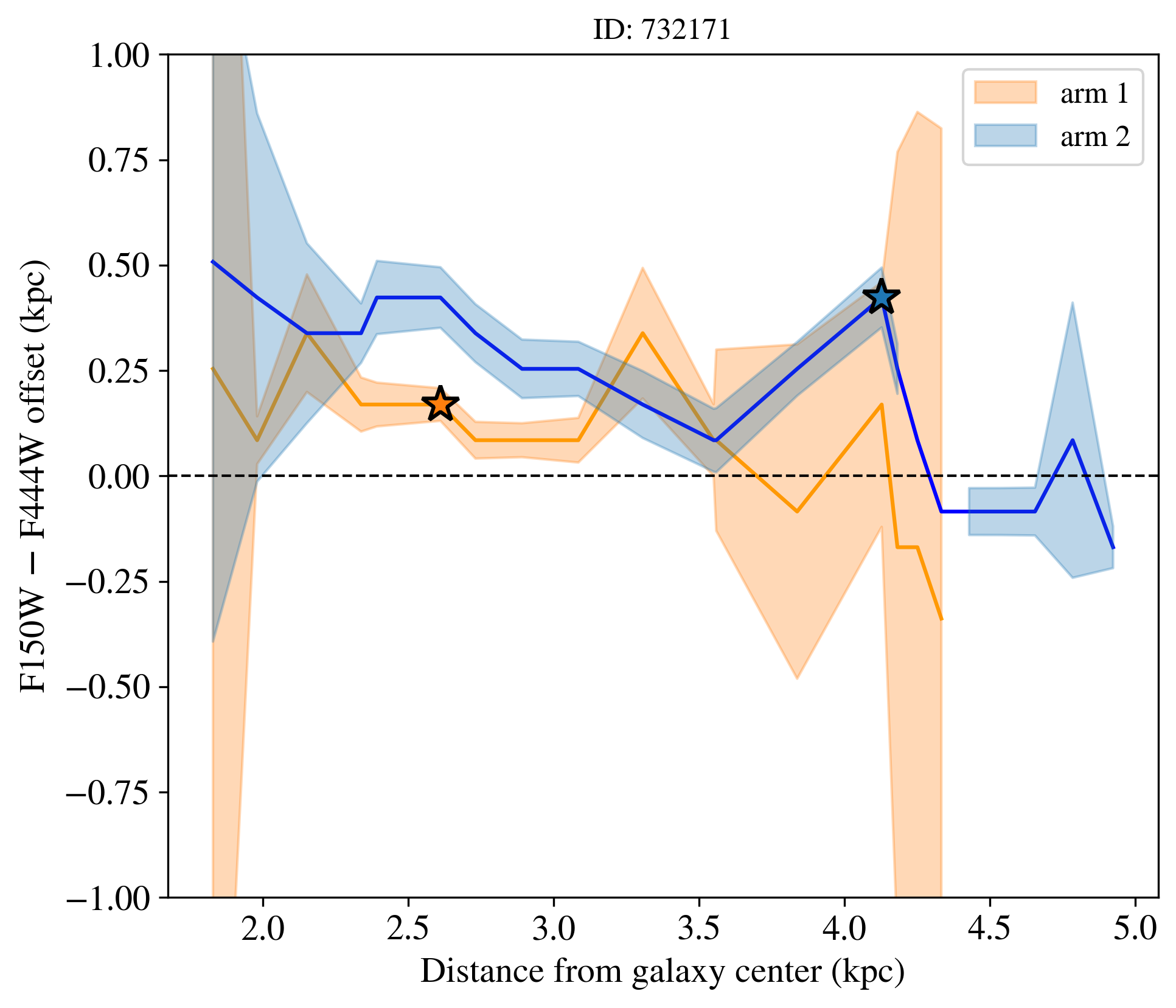}
    \includegraphics[width=0.45\textwidth]{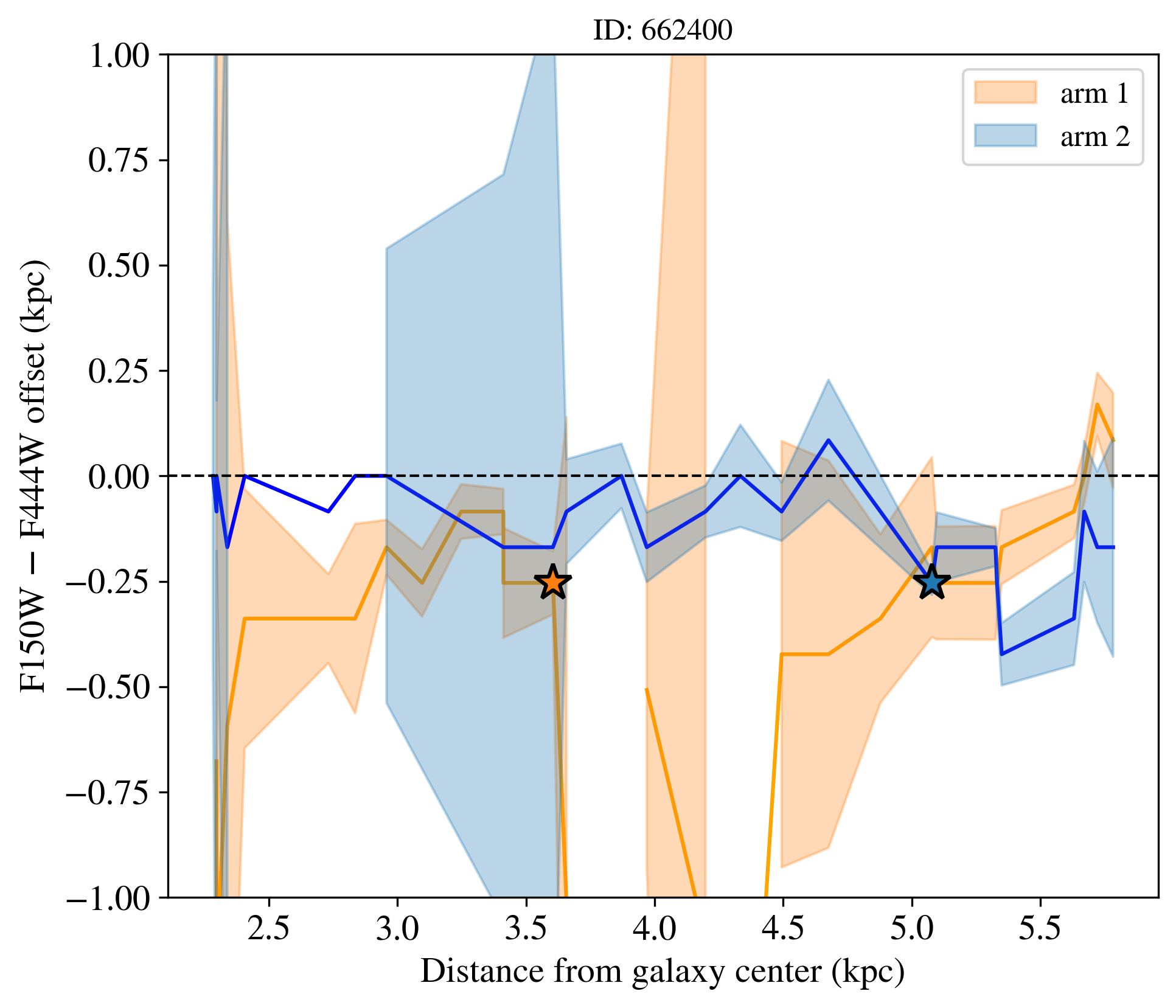}
    \caption{Radial distribution of offsets: The radial variation of the offset between the peaks of the F150W and F444W flux distributions across the spiral arms for the galaxies shown in Fig.~\ref{fig:pos_offset}: ids 732171 (left) and 662400 (right).  The corresponding shaded regions indicate the $1\sigma$ uncertainties. The stars show the datapoint taken as the maximum offset value, determined using the maximum absolute value and the corresponding uncertainty.}
    \label{fig:offset_dist}
\end{figure*}

\begin{figure} 
    \centering
    \includegraphics[width=0.45\textwidth]{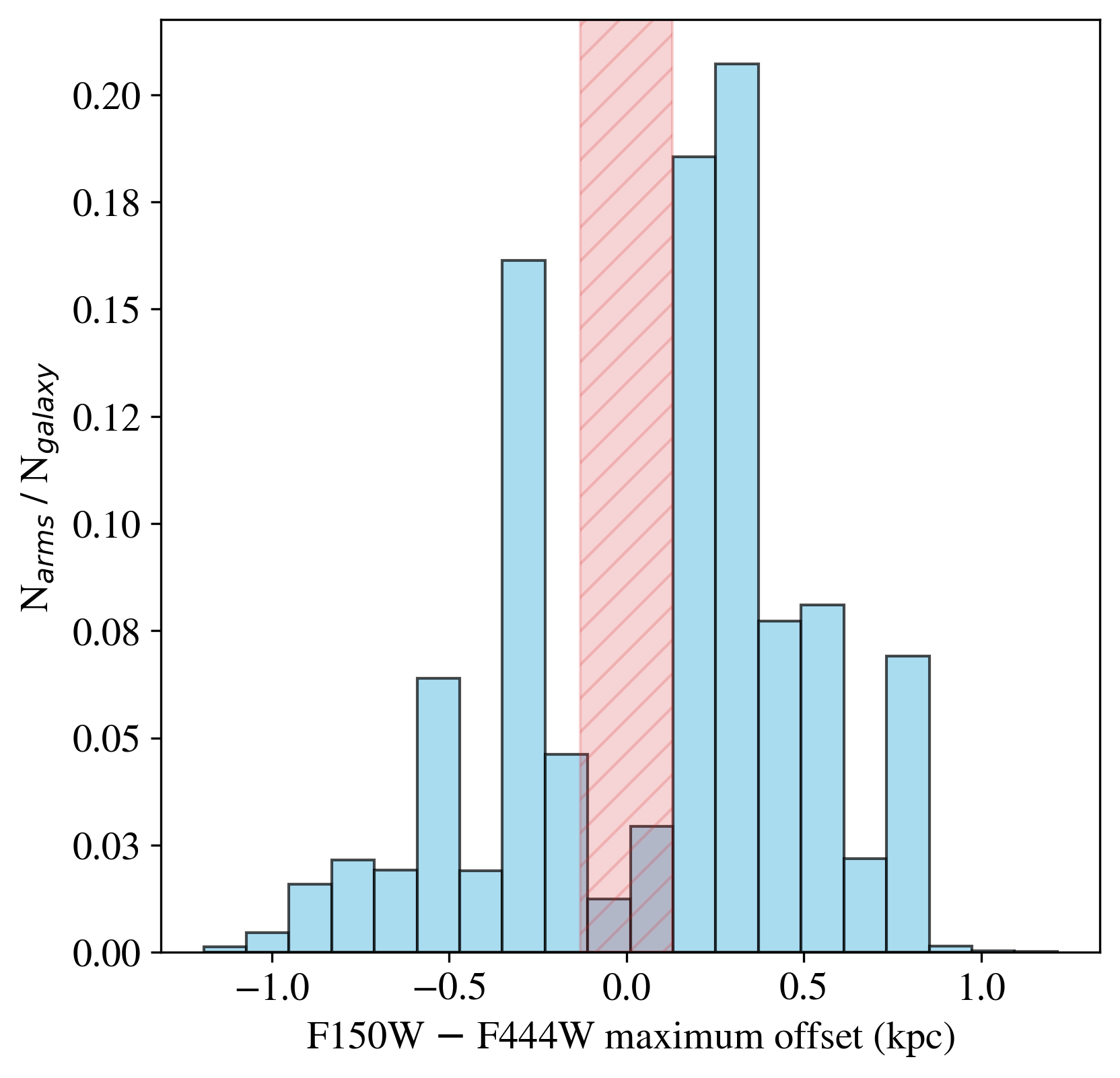}
    \caption{Maximum offset distribution: The normalized number distribution of the maximum offset between the F150W and F444W flux distributions across the spiral arms in our sample. For our sample, $\rm N_{\rm galaxy} = 8$. The x-axis is shown in  kpc (for $z=1.5$). We create this histogram by randomly sampling values (1000 times) within the $1 \sigma$ error margins for each maximum offset measurement. The total uncertainty of our method is estimated to be $\sim 0.13\,\rm kpc$, so values within $0 \pm 0.13\,\rm kpc$ (region in red) are considered to indicate no offset.}
    \label{fig:offset_hist}
\end{figure}

Within the sample of 8 galaxies, we detect and model 18 spiral arms (Fig.~\ref{fig:spiral_analysis}). For each arm, we measure the flux variation at each iteration step.  15/18 arms show a peak F444W flux $\geq 5\sigma$ over $60$--$100\%$ of their lengths (contours shown in Fig.~\ref{fig:image_comp}, middle panel). For the remaining three, the fraction of length being detected drops to $\sim 10$--$30\%$.  In contrast, the `well-detected' 15/18 arms have a peak F150W flux $\geq 5\sigma$ over $\sim 10$--$80\%$ of their lengths. However, considering the F150W image is 0.5 mag shallower than F444W, a more appropriate threshold is $\sim 3\sigma$, where arms are detected over $20$--$90\%$ of their lengths (contours shown in Fig.~\ref{fig:image_comp}, bottom panel). The final three arms, that had lower F444W detections, are not detected in the F150W data.  Given that the flux per unit frequency ($\rm F_{\nu}$) should remain almost constant over the rest-frame wavelength range covered by F150W and F444W at $z = 1.5$ for a constant star-formation model without dust attenuation, there is clear reddening in the arms, likely caused by dust or stellar age. Since spiral arms are predominantly star-forming, attenuation likely plays a major role.   The three arms that were weakly detected in F444W,  were likely pushed below our F150W detection thresholds due to attenuation. 

We also find clear spatial offsets between the flux distributions in F444W and F150W based on their flux peaks (Fig.~\ref{fig:pos_offset}). Additionally, we assign a direction to the spiral arms based on the expected propagation for density waves (Fig.~\ref{fig:pos_offset}). A positive offset indicates that the near-IR flux lags behind the optical flux, while a negative offset shows the reverse. We observe that the offset is not constant along the arms but shows clear radial variations (Fig.~\ref{fig:offset_dist}). Averaging over the length of the spiral arms, the offsets for the 15/18 arms with both F444W and F150W detections are mostly within $-0.2\,\rm kpc$ to $0.4\,\rm kpc$, with one exception at $-0.4\,\rm kpc$. Given our spatial uncertainty of $0.13\,\rm kpc$, this suggests a marginal bias toward positive values. 

However, given the radial variation, the average is not a robust tracer of the offsets. We rather determine the maximum offset for each arm by finding the location where the following parameter ($F$) is maximised:
\begin{equation}
    F = \alpha\,|\rm offset| + (1-\alpha)\,significance
\end{equation}
Here, the `offset' is the difference between the F150W and F444W flux peaks. The `significance' refers to the ratio of the offset and the corresponding error. This allows us to determine the maximum offset not just based on the absolute value but also the significance. We choose $\alpha = 0.5$ to allow an equal contribution from both these parameters. The radial locations of the maximum offsets for arms in two of the galaxies in our sample (the same as in Fig.~\ref{fig:pos_offset}) are provided in Fig.~\ref{fig:offset_dist}. 10/15 detected arms show positive offsets, with nine of these showing offsets greater than our systematic uncertainty ($\sim 0.13\,\rm kpc$, Sec.~\ref{subsec:mapping_flux}) ranging from $0.2\,\rm kpc$ to $0.8\,\rm kpc$ (Fig.~\ref{fig:offset_hist}). Meanwhile, 5/15 arms show negative values from $-0.2\,\rm kpc$ to $-0.8\,\rm kpc$, with three of these arms found in two galaxies (ids 449617 and 662400) showing clear signs of interaction.

\section{Discussions} \label{sec:discussion}

In this section, we shall be proposing that the observed offset between the flux peaks in rest-frame near-IR and optical wavelengths indicates density wave propagation, either quasi-static or transient. For long-lived quasi-static spirals, the wave-like arms have fixed angular velocities. Thus, below the co-rotation radius, the arms propagate slower than the differentially rotating disk (trailing arms). However, in the case of transient waves, deviations from a fixed angular velocity are expected. Nevertheless, we generally still expect trailing arms \citep{dobbs14}.

As gas in the disk falls into the potential of these trailing arms, it can be accelerated to the speed of sound, creating shocks. The initial `spiral-shock' would therefore lag behind the arm \citep{Roberts1969}. The compression of the gas clouds causes part of the gas to lose angular momentum and flow inward \citep{Kim14}, while some of it forms new stars and moves through the density wave \citep{Roberts1969, gittins04}. The majority of the dust produced in this process, tracing the location of the shock, resides on the leading side of the rest-frame near-IR arm \citep{yu18}. As a result, the optical flux from the newly formed young stars is attenuated, while the young stars that have moved past the slowly propagating arm experience less attenuation. This combination produces an optical wavelength flux concentration in front of the arm. 

The positive offsets we observe in half of the spiral arms in our sample match these expectations (Figs.~\ref{fig:pos_offset} and \ref{fig:offset_hist}), assuming the arms are predominantly below the co-rotation radius.  Furthermore, in the case of quasi-static density waves, the positive offset should gradually decrease with increasing radius as the co-rotation radius is approached.  The lowering of the positive offset in galaxy ID: 732171 at the highest radial distances for both of its spiral arms (Fig.~\ref{fig:offset_dist}, left panel) could be interpreted as a sign of such a gradient. However, proper quantification of this effect will require larger statistics and kinematic information.

Within the density wave scenario, the offset magnitude will be influenced by time lag between gas compression and star formation, the dust distribution and the velocity offset between the arm and the disk. Disentangling these factors will be crucial to understanding the dynamics of the spiral arm relative to the underlying disk. Deeper sub-mm images tracing dust emission and spectroscopic data providing the velocities of different disk components will be necessary. Nevertheless, the observed red-to-blue gradient aligns well with studies on density wave propagation at low redshifts \citep[$0 \lesssim z \lesssim 1$][]{yu18,Martinez-Garcia2023}. These studies also predict a secondary blue-to-red gradient from aging stars moving away from the arms, which future studies should investigate.

Meanwhile, five arms show a clear negative offset (e.g., Fig.~\ref{fig:pos_offset}, right panel).  It is worth noting that none of the respective host galaxies simultaneously feature arms with positive offsets.  Explaining this through density waves requires invoking motion beyond the co-rotation radius while simultaneously observing positive offsets at lower radii, which we do not observe. Instead, some local cases show waves from interactions causing arms to move faster than the disk medium \citep{buta92, buta03}. We may be observing a similar effect here, especially since minor-mergers increase with redshift \citep{lotz11}, making them common at $z \sim 1.5$. In this scenario, the spiral arms are density waves accelerated by the external potential of a companion. Three of the five arms with a negative offset appear in the two interacting galaxies in our sample, supporting this conclusion. 

Finally, three arms are not detected in rest-frame optical (F150W) while being marginally detected in rest-frame near-IR (F444W). These cases likely involve formation mechanisms without velocity offsets, such as interaction-driven material arms or stochastic star formation.  The latter is hypothesized to produce weak flocculent spirals \citep{gerola78,jungweirt94,sleath95}, which would be difficult to model.  Stochastic star formation is also important because galaxies at $z>1$ are highly clumpy \citep[e.g.,][]{elmegreen08,genzel11,forster11,claeyssens23,sattari23,kalita24,faisst24}, driven by high gas fractions \citep{daddi10,tacconi10,geach11,rujopakarn23}, leading to large fractions of star formation (up to $\sim 25\%$) concentrated in kpc-scale complexes.  The resulting massive clumps can be sheared by the rotating disk, leading to spiral structures \citep{elmegreen14,genzel23}.  We also expect to be underestimating its impact in our study, having excluded galaxies with faint, patchy arms -- key traits of flocculent spirals.  Hence, spiral arms in such galaxies would probably appear in the region between the positive and negative peaks in Fig.~\ref{fig:offset_hist}. Furthermore, feedback from star formation may also erase key signatures of density-wave spirals \citep{shetty08,dobbs11}, complicating assessments of formation mechanisms.



\section{Summary} \label{sec:summary}
We use JWST/NIRCam data for eight massive star-forming galaxies (stellar mass $ = 10^{10.5-11.4}\,\rm M_{\odot}$) at $z_{\rm spec}\sim1.5$ to characterize their spiral arms in detail. Using Cartesian and polar shapelet transforms, we isolate the spiral structures in rest-frame near-IR and construct models. These models chart the radial paths along the spirals, allowing us to iteratively measure the flux distribution along these paths and across the width of the arms in PSF-matched rest-frame optical (F150W NIRCam filter) and near-IR (F444W NIRCam filter) residual images.

We find that all eighteen spiral arms in our sample are detected at $\geq 5\sigma$ significance in the stellar mass-tracing near-IR wavelength. However, 3/18 are entirely absent in rest-frame optical, and in the remaining arms, the detected length fraction significantly decreases compared to the near-IR, likely due to attenuation by dust known to exist in spiral arms. Half of the arms show robust levels of red-to-blue color gradient along the expected direction of arm propagation. Drawing parallels to the local Universe, we interpret this gradient as an indicator of dust from shocks caused by density wave-driven spiral propagation, where the arms are slower than the disk's circular velocity. Conversely, five arms display a blue-to-red gradient, likely due to tidal interaction perturbations. Arms without any gradient are probably driven by interactions or stochastic star formation. We also predict that spirals formed through stochastic star-formation may be under-represented in our study due to their patchy nature, making them challenging to model.

In conclusion, this study provides a quantitative characterization of spiral structures in the $z > 1$ Universe, made possible by JWST. We highlight the complex, multi-faceted nature of spiral arms and emphasize the need for further studies to understand galaxy morphological evolution during this critical epoch. 


\begin{acknowledgements}
We thank the referee for the valuable comments.  B.S.K. and J.D.S. are supported by the World Premier International Research Center Initiative (WPI), MEXT, Japan. J.D.S. is supported by JSPS KAKENHI (JP22H01262). This work was also supported by JSPS Core-to-Core Program (grant number: JPJSCCA20210003). O.I. acknowledges the funding of the French Agence Nationale de la Recherche for the project iMAGE (grant ANR-22-CE31-0007).  L.C.H. was supported by the National Science Foundation of China (11991052, 12233001), the National Key R\&D Program of China (2022YFF0503401), and the China Manned Space Project (CMS-CSST-2021-A04, CMS-CSST-2021-A06). This work was made possible by utilising the CANDIDE cluster at the Institut d’Astrophysique de Paris. The cluster was funded through grants from the PNCG, CNES, DIM-ACAV, the Euclid Consortium, and the Danish National Research Foundation Cosmic Dawn Center (DNRF140). It is maintained by Stephane Rouberol.  The JWST data presented in this article were obtained from the Mikulski Archive for Space Telescopes (MAST) at the Space Telescope Science Institute. The specific observations analyzed can be accessed via \dataset[doi: 10.17909/04ww-nn64]{https://doi.org/10.17909/04ww-nn64}. The dataset was obtained as part of the COSMOS-Web survey program (id 1727).  We would finally like to thank Pascal Oesch and Mengyuan Xiao for the valuable discussions. 
\end{acknowledgements}



\bibliography{main}{}
\bibliographystyle{aasjournal}


\end{document}